\documentclass[a4paper,11pt]{article}
\usepackage[left=3.17cm,right=3.17cm,top=2.54cm,
headheight=0.5cm,headsep=0.54cm,bottom=2.54cm,footskip=0.79cm
]{geometry}

\usepackage{amsmath,amssymb}

\usepackage{tikz}
\usetikzlibrary{arrows}
\usetikzlibrary{decorations.markings}

\usepackage[pdfstartview=FitH]{hyperref}

\begin{document}

\title{A groupoidification of the fermion algebra}

\author{ Wei Chen, Bing-Sheng Lin$\thanks{e-mail: sclbs@scut.edu.cn}$\\[.3cm]
{\small School of Mathematics, South China University of Technology,
Guangzhou 510641, China.}
}

\maketitle

\begin{abstract}
In this paper, we consider the groupoidification of the fermion algebra. We construct a groupoid as the categorical analogues of the fermionic Fock space, and the creation and annihilation operators correspond to spans of groupoids. The categorical fermionic Fock states have some extra structures comparing with the normal forms. We also construct a 2-category of spans of groupoids corresponding to the fermion algebra. The relations of the morphisms in this 2-category are consistent with those in the graphical category which is represented by string diagrams.
\\


\end{abstract}

\section{Introduction}\label{sec1}
In recent years, there has been much interest in the studies of categorifications of theories in mathematics and theoretical physics \cite{Baez}-\cite{cl}. In general, categorification is a process of replacing set-theoretic theorems by category-theoretic analogues. It replaces sets by categories, functions by functors, and equations between functions by natural transformations of functors \cite{Baez}.
Categorification can be thought of as the process of enhancing an algebraic object to a more sophisticated one, while ¡°decategorification¡± is the process of reducing the categorified object back to the simpler original object. So a useful categorification should possess a richer structure not seen in the underlying object. The categorification of physical theories may extend the mathematical
structures of existing physical theories and help us solve the remaining problems in fundamental physics, it can also help us better understand the physical essence.

Groupoidification is a form of categorification in which vector spaces are replaced by groupoids and linear operators are replaced by spans
of groupoids \cite{Groupoid}. A groupoid is a special type of category in which every morphism is invertible. It can be seen as a generalization of a group, and a usual group is a groupoid where there is only one object.
In the framework of groupoidification, the configuration spaces for a physical system are described by some groupoids, and the physical histories are described by spans of groupoids. Furthermore, one can encode the symmetries of the physical system in arrows in the groupoids.

In Ref.~\cite{Groupoid}, the authors considered the groupoidification of the quantum harmonic oscillator system, and the Hilbert space for the quantum harmonic oscillator arises naturally from degroupoidifying the groupoid of finite sets and bijections, and these give a purely combinatorial interpretation of creation and annihilation operators. The authors also considered the groupoidification of the field operators, their normal-ordered powers and the corresponding Feynman diagrams.
In Ref.~\cite{CHA}, the authors also constructed a 2-category of spans of groupoids for the Heisenberg algebra, and gave a combinatorial model for Khovanov's diagrammatic categorification of the Heisenberg algebra \cite{Khovanov}.

Since the Heisenberg algebra (also called boson algebra) and the fermion algebra are the most fundamental algebraic relations in quantum physics, it is significant to study the groupoidification of the fermion algebra. In the present paper, by virtue of the methods developed in Refs.~\cite{Groupoid, CHA}, we construct a groupoid corresponding to the fermionic Fock space, and the creation and annihilation operators correspond to some type of spans of groupoids.

This paper is organized as follows. In Section \ref{sec2}, we will briefly review the properties of the 1D fermion algebra in normal quantum mechanics. In Section \ref{sec3}, we review the main results of the diagrammatic categorification of the fermion algebra in Ref.~\cite{lwwy}. In Section \ref{sec4}, we study the groupoidification of the fermion algebra, and compare the results with the diagrammatic categorification of the fermion algebra using string diagrams.
Some discussions are given in Section \ref{sec5}.

\section{The fermion algebra}\label{sec2}
In the following contents, we will only consider the one-dimensional fermion algebra, and the results can be easily extended to those of higher-dimensional fermion algebras.
In normal quantum mechanics, the fermionic creation and annihilation operators $\hat{f}^{\dag}$, $\hat{f}$ satisfy the fermion algebraic relations
\begin{eqnarray}\label{1df}
&&\{\hat{f},\hat{f}^{\dag}\}:=\hat{f} \hat{f}^{\dag}+\hat{f}^{\dag} \hat{f}=1,\nonumber\\
&&\{\hat{f},\hat{f}\}=\{\hat{f}^{\dag},\hat{f}^{\dag}\}=0.
\end{eqnarray}
Obviously, we have $\hat{f} \hat{f}=0$ and $\hat{f}^{\dag}\hat{f}^{\dag}=0$.

The corresponding Hilbert space is spanned only by two states, which can be denoted by $|0\rangle$ and $|1\rangle$. These are single mode fermionic Fock states, and satisfy the following relations
\begin{eqnarray}\label{fs}
&&\hat{f}|0\rangle=0,\qquad \hat{f}^{\dag}|0\rangle =|1\rangle, \nonumber\\
&&\hat{f}|1\rangle =|0\rangle,\qquad \hat{f}^{\dag}|1\rangle =0.
\end{eqnarray}
The states $|0\rangle$ and $|1\rangle$ are orthonormal,
\begin{equation}\label{orth}
\langle 0|0\rangle=\langle 1|1\rangle=1,
\qquad\langle 0|1\rangle=\langle 1|0\rangle=0.
\end{equation}

We also have the following matrix representations
\begin{equation}\label{mat}
\hat{f}=\left(
           \begin{array}{cc}
             0 & 0 \\
             1 & 0 \\
           \end{array}
         \right),\qquad
\hat{f}^{\dag}=\left(
           \begin{array}{cc}
             0 & 1 \\
             0 & 0 \\
           \end{array}
         \right),\qquad
|0\rangle=\left(
           \begin{array}{cc}
             0 \\
             1 \\
           \end{array}
         \right),\qquad
|1\rangle =\left(
           \begin{array}{cc}
             1 \\
             0 \\
           \end{array}
         \right).
\end{equation}

\section{Diagrammatic categorification of the fermion algebra}\label{sec3}
In Ref.~\cite{lwwy}, the authors constructed a diagrammatic categorification of the fermion algebra, here we will briefly review the main results.
Let $\mathcal{F}$ be an additive $\Bbbk$-linear strict monoidal category for a commutative ring $\Bbbk$, and the set of objects in $\mathcal{F}$ is generated by objects $Q_+$ and $Q_-$. An arbitrary object of $\mathcal{F}$ is a finite direct sum of tensor products $Q_\varepsilon := Q_{\varepsilon_1} \otimes
\dots \otimes Q_{\varepsilon_n} $, and $\varepsilon = \varepsilon_1
\dots \varepsilon_n$ is a finite sequence of $+$ and $-$ signs. The unit object is $\mathbf{1}=Q_\emptyset$. The objects $Q_+$ and $Q_-$ can be regarded as the categorical analogues of the fermionic creation and annihilation operators $\hat{f}^{\dag}$, $\hat{f}$.

The morphisms in $\mathcal{F}$ are denoted by string diagrams. The diagrams are
oriented compact one-manifolds immersed in the strip $\mathbb{R} \times
[0,1]$, modulo rel boundary isotopies.  The endpoints of the
one-manifold are located at $\{1,\dots,m\} \times \{0\}$ and
$\{1,\dots,k\} \times \{1\}$, where $m$ and $k$ are the lengths of
the sequences $\varepsilon$ and $\varepsilon'$ respectively.  The
orientation of the one-manifold at the endpoints must agree with the
signs in the sequences $\varepsilon$ and $\varepsilon'$.
For example, the diagram
\begin{equation}
\begin{tikzpicture}[>=angle 60,thick,baseline=0pt]
  \draw[<-] (0,.5) to (1,-.5);
  \draw[<-] (0,-.5) to (1,.5);
\end{tikzpicture}
\end{equation}
is one of the morphisms from $Q_{-+}$ to $Q_{+-}$. A diagram without endpoints gives an endomorphism of $\mathbf{1}$.

The space of morphisms $\mathrm{Hom}_{\mathcal{F}}(Q_\varepsilon,
Q_{\varepsilon'})$ is the $\Bbbk$-module generated by string diagrams
modulo local relations.
The local relations for the morphisms in $\mathcal{F}$ are as follows.

\begin{equation}\label{lcr1}
\begin{tikzpicture}[>=angle 60,thick,baseline=0pt]
  \useasboundingbox (-.2,-1.1) rectangle (7.1,1.1);
  \draw (0,.75) arc (180:360:.5) ;
  \draw (0,1) -- (0,.75) ;
  \draw (1,1) -- (1,.75) [<-];
  \draw (1,-.75) arc (0:180:.5) ;
  \draw (1,-1) -- (1,-.75) ;
  \draw (0,-1) -- (0,-.75) [<-];
  \draw (1.5,0) node {$=$};
  \draw (2,-1) -- (2,1) [<-];
  \draw (3,-1) -- (3,1) [->];
  \draw (3.5,0) node {,};
  \draw (5,-1) -- (5,1) [->];
  \draw (6,-1) -- (6,1) [->];
  \draw (6.5,0) node {$=$};
  \draw (7,0) node {$0$};
\end{tikzpicture}
\end{equation}
\ \\
\begin{equation}\label{lcr2}
\begin{tikzpicture}[>=angle 60,thick,baseline=0pt,
decoration={markings,mark=at position 0 with {\arrowreversed[]{angle 60}}}]
  \draw [postaction={decorate}] (.5,0) circle (.5);
  \draw (1.5,0) node {$+$};
  \draw[xscale=-1,shift={(-5,0)}] [postaction={decorate}] (2.5,0) circle (.5);
  \draw (3.5,0) node {$=$};
  \draw (4,0) node {$\mathrm{id}$};
  \useasboundingbox (-.1,-.6) rectangle (4.1,.6);
\end{tikzpicture}
\end{equation}
\ \\
The relations (\ref{lcr1}) are equivalent to the following relations
\begin{equation}\label{lcr3}
\begin{tikzpicture}[>=angle 60,thick,baseline=0pt]
  \useasboundingbox (-.2,-1.1) rectangle (7.1,1.1);
  \draw (0,.75) arc (180:360:.5) ;
  \draw (0,1) -- (0,.75) [<-];
  \draw (1,1) -- (1,.75);
  \draw (1,-.75) arc (0:180:.5) ;
  \draw (1,-1) -- (1,-.75) [<-];
  \draw (0,-1) -- (0,-.75);
  \draw (1.5,0) node {$=$};
  \draw (2,-1) -- (2,1) [->];
  \draw (3,-1) -- (3,1) [<-];
  \draw (3.5,0) node {,};
  \draw (5,-1) -- (5,1) [<-];
  \draw (6,-1) -- (6,1) [<-];
  \draw (6.5,0) node {$=$};
  \draw (7,0) node {$0$};
\end{tikzpicture}
\end{equation}
The second relation in (\ref{lcr1}) means that $Q_{++}\cong \mathbf{0}$, where $\mathbf{0}$ is zero object in the additive category $\mathcal{F}$. There is also $Q_{--}\cong\mathbf{0}$. These isomorphic relations just correspond to the operator relations $\hat{f}^{\dag}\hat{f}^{\dag}=0$ and $\hat{f}\hat{f}=0$.
Furthermore, one may find that all the string diagrams with crossings are equal to zero.

From the local relations above, one may obtain the following isomorphic relation in the category $\mathcal{F}$
\begin{equation}\label{fa}
  Q_{-+} \oplus Q_{+-} \cong \mathbf{1},
\end{equation}
so in the Grothendieck group $K_0(\mathcal{F})$, we have
\begin{equation}
[Q_-][Q_+] +[Q_+][Q_-] = 1,
\end{equation}
which is the fermion algebraic relation (\ref{1df}).

\section{Groupoidification of the fermion algebra}\label{sec4}
In the framework of groupoidification, groupoids and spans can be degroupoidified to vectors and linear operators, which generally can be represented by matrices. The groupoids and spans also have natural physical meanings \cite{Groupoid}.
For example, let us consider the following span of sets
\begin{equation}
\begin{tikzpicture}[>=angle 60,thick,baseline=0pt]
  \node (a) at (0,0.8) {$M$};
  \node (b) at (2,-.8) {$A$};
  \node (c) at (-2,-.8) {$B$};
  \draw[->] (a) to node [above left] {$g$} (c);
  \draw[->] (a) to node [above right] {$f$} (b);
\end{tikzpicture}
\end{equation}
$A$ can be considered as a set whose elements are possible initial states for the physical system, and $B$ is the set whose elements are possible final states, then $M$ can be considered as a set of possible events, or histories. For example, let $i\in A$ denotes the $i$th initial state and $j\in B$ denotes the $j$th final state, then the following subset of $M$
\begin{equation}
M_{ji}=\{m: f(m)=i, g(m)=j\}
\end{equation}
can be regarded as the set of ways for the physical system to undergo a transition from its $i$th initial state to the $j$th final state.
If all the sets $M_{ij}$ are finite, then the $(i,j)$-elements of the correspond matrix under degroupoidification are just the cardinalities $|M_{ij}|$.
When $A$, $B$ and $M$ are groupoids, the objects of $A$ and $B$ can be regarded as the corresponding physical states. Furthermore, in the groupoids, one can also consider the morphisms between the objects, these morphisms can represent some symmetries of the physical system.

Two spans $(B\xleftarrow{g}M\xrightarrow{f}A)$ and $(B\xleftarrow{g'}N\xrightarrow{f'}A)$ are isomorphic if there is an isomorphism $h: M\to N$ satisfying the following commuting diagram \cite{2Vec}
\begin{equation}\label{iso}
\begin{tikzpicture}[>=angle 60,thick,baseline=0pt]
  \node (a) at (0,0.8) {$M$};
  \node (b) at (2,-.8) {$A$};
  \node (c) at (-2,-.8) {$B$};
  \node (d) at (0,-.8) {$N$};
  \draw[->] (a) to node [above left] {$g$} (c);
  \draw[->] (a) to node [above right] {$f$} (b);
  \draw[->] (d) to node [below] {$g'$} (c);
  \draw[->] (d) to node [below] {$f'$} (b);
  \draw[->] (a) to node [right] {$h$} (d);
\end{tikzpicture}
\end{equation}

In Ref.~\cite{CHA}, the authors considered the groupoidification of the Heisenberg algebra. They categorified the bosonic Fock state $|n\rangle$ by the $n$-element set. This is intuitive and reasonable, since the Fock state $|n\rangle$ means there are $n$ quanta or particles.
The bosonic Fock space is naturally represented by the groupoid $\mathbf{S}$ of finite sets and bijections.
The creation and annihilation operators $\hat{a}^{\dag}$ and $\hat{a}$ are represented by the following spans of groupoids $\mathbf{S}$,
\begin{equation}\label{ff}
\begin{tikzpicture}[>=angle 60,thick,baseline=0pt]
  \node (a) at (0,0.8) {$\mathbf{S}$};
  \node (b) at (2,-.8) {$\mathbf{S}$};
  \node (c) at (-2,-.8) {$\mathbf{S}$};
  \draw[->] (a) to node [above left] {+1} (c);
  \draw[->] (a) to node [above right] {$\mathrm{id}$} (b);
  \draw (0,-1.6) node {$\hat{a}^{\dag}$};
\end{tikzpicture}\qquad\qquad
\begin{tikzpicture}[>=angle 60,thick,baseline=0pt]
  \node (a) at (0,.8) {$\mathbf{S}$};
  \node (b) at (-2,-.8) {$\mathbf{S}$};
  \node (c) at (2,-.8) {$\mathbf{S}$};
  \draw[->] (a) to node [above left] {$\mathrm{id}$} (b);
  \draw[->] (a) to node [above right] {+1} (c);
  \draw (0,-1.6) node {$\hat{a}$};
\end{tikzpicture}
\end{equation}
and $\mathbf{S}\xrightarrow{+1}\mathbf{S}$ is the functor taking the disjoint union with the one-element set.

For the fermion algebra, the corresponding Fock space is spanned only by two states, namely $|0\rangle$ and $|1\rangle$. There are no any nontrivial combinatorial models corresponding to these states, so we can not construct the categorical analogues of the fermionic Fock states with the constructions used in Ref.~\cite{CHA}.

Intuitively, there is some duality between the Fock states $|0\rangle$ and $|1\rangle$ from the relations (\ref{fs}), so one may consider the categorical analogues of the states $|0\rangle$ and $|1\rangle$ as some object $A$ and its dual object $A^*$ in some category (e.g., a compact closed category). For example, in the category $\mathbf{FdVect}$ with finite-dimensional vector spaces as objects and linear maps as morphisms, $A$ is a finite-dimensional vector space, and $A^*$ is its dual space.

Let $\mathbf{\Psi}$ be a compact closed category containing an object $A$ and its dual object $A^*$. All the morphisms in $\mathrm{Hom}_{\mathbf{\Psi}}(A,A)$ and $\mathrm{Hom}_{\mathbf{\Psi}}(A^*,A^*)$ are invertible, namely automorphisms, and $\mathrm{Hom}_{\mathbf{\Psi}}(A,A^*)=\mathrm{Hom}_{\mathbf{\Psi}}(A^*,A)=\varnothing$.
In fact, in this construction, the morphisms $\mathrm{Hom}_{\mathbf{\Psi}}(A,A)$ form a group, namely the automorphism group of $A$, which can be denoted by $\mathrm{Aut}_{\mathbf{\Psi}}(A)$, or simply $\mathrm{Aut}(A)$. Similarly, there is an automorphism group $\mathrm{Aut}(A^*)$ of the object $A^*$. Furthermore, one may naturally assume $\mathrm{Aut}(A)\cong\mathrm{Aut}(A^*)$.
$\mathbf{\Psi}$ is also a groupoid, it can be regarded as a groupoidification of the fermionic Fock space, and the objects $A$ and $A^*$ correspond to the states $|0\rangle$ and $|1\rangle$, respectively.

We may construct the categorical analogues of the fermionic creation and annihilation operators $\hat{f}^{\dag}$, $\hat{f}$ with the aid of spans of groupoids. Let us define the spans of groupoids as follows,
\begin{equation}
\begin{tikzpicture}[>=angle 60,thick,baseline=0pt]
  \node (a) at (0,0.8) {$\mathbf{H}$};
  \node (b) at (2,-.8) {$\mathbf{\Psi}$};
  \node (c) at (-2,-.8) {$\mathbf{\Psi}$};
  \draw[->] (a) to node [above left] {$T$} (c);
  \draw[->] (a) to node [above right] {$I$} (b);
  \draw (0,-1.6) node {$F^{\dag}$};
\end{tikzpicture}\qquad\qquad
\begin{tikzpicture}[>=angle 60,thick,baseline=0pt]
  \node (a) at (0,.8) {$\mathbf{H}$};
  \node (b) at (-2,-.8) {$\mathbf{\Psi}$};
  \node (c) at (2,-.8) {$\mathbf{\Psi}$};
  \draw[->] (a) to node [above left] {$I$} (b);
  \draw[->] (a) to node [above right] {$T$} (c);
  \draw (0,-1.6) node {$F$};
\end{tikzpicture}
\end{equation}
Here $\mathbf{H}$ is a full subcategory of $\mathbf{\Psi}$ containing only one object $A$, and $\mathrm{Hom}_{\mathbf{H}}(A,A)=\mathrm{Hom}_{\mathbf{\Psi}}(A,A)$. Obviously, $\mathbf{H}$ is also a groupoid, in fact, a group.
The inclusion functor $I: \mathbf{H}\to\mathbf{\Psi}$ takes objects and morphisms to themselves, and $T$ is a contravariant functor takes the object $A\mapsto A^*$ and morphisms $\mathrm{Hom}_{\mathbf{H}}(A,A)\ni f\mapsto f^*\in\mathrm{Hom}_{\mathbf{\Psi}}(A^*,A^*)$. We denote these spans by $F^{\dag}$ and $F$, respectively.
These spans are the categorical analogues of the fermionic creation and annihilation operators $\hat{f}^{\dag}$ and $\hat{f}$.

Similar to Ref.~\cite{CHA}, one may construct a 2-category $\mathcal{C}$ of spans of groupoids corresponding to the graphical category $\mathcal{F}$ in the previous section.
In $\mathcal{C}$, objects are some tame groupoids, 1-morphisms are isomorphism classes of spans of groupoids, with composition defined by weak pullback, and 2-morphisms are isomorphism classes of spans of spans.
So $\mathbf{\Psi}$ is an object in $\mathcal{C}$, and the spans of groupoids $F^{\dag}$ and $F$ are just 1-morphisms in $\mathcal{C}$.

Now let us consider the composition of 1-morphisms in $\mathcal{C}$, which are just composition of the corresponding spans. Here we will use the notations used in Ref.~\cite{CHA}.
For example, the composition of the spans of groupoids $(B\xleftarrow{G}X\xrightarrow{H}A)$ and $(C\xleftarrow{K}Y\xrightarrow{J}B)$ is the following span
\begin{equation}
\begin{tikzpicture}[>=angle 60,thick,baseline=0pt]
  \node (a) at (0,1.6) {$(J\downarrow G)$};
  \node (b) at (-2,0) {$Y$};
  \node (c) at (2,0) {$X$};
  \node (x) at (-4,-1.6) {$C$};
  \node (y) at (0,-1.6) {$B$};
  \node (z) at (4,-1.6) {$A$};
  \draw[->] (a) to node [above left] {$P_Y$} (b);
  \draw[->] (a) to node [above right] {$P_X$} (c);
  \draw[->] (b) to node [above left] {$K$} (x);
  \draw[->] (b) to node [above right] {$J$} (y);
  \draw[->] (c) to node [above left] {$G$} (y);
  \draw[->] (c) to node [above right] {$H$} (z);
\end{tikzpicture}
\end{equation}
$(J\downarrow G)$ is a weak pullback groupoid where an object is a triple $(x,y,f)$ consisting of an object $x\in \mathrm{Ob}(X)$, an object $y\in \mathrm{Ob}(Y)$, and an isomorphism $f:G(x)\to J(y)$ in $B$. A morphism $(x_1,y_1,f_1)\to(x_2,y_2,f_2)$ in $(J\downarrow G)$ consists of morphisms $x_1\xrightarrow{a}x_2$ and $y_1\xrightarrow{b}y_2$ satisfying the following commuting diagram
\begin{equation}
\begin{tikzpicture}[>=angle 60,thick,baseline=0pt]
  \node (a) at (-1,1) {$G(x_1)$};
  \node (b) at (-1,-1) {$G(x_2)$};
  \node (c) at (1,1) {$J(y_1)$};
  \node (d) at (1,-1) {$J(y_2)$};
  \draw[->] (a) to node [left] {$G(a)$} (b);
  \draw[->] (a) to node [above] {$f_1$} (c);
  \draw[->] (b) to node [below] {$f_2$} (d);
  \draw[->] (c) to node [right] {$J(b)$} (d);
\end{tikzpicture}
\end{equation}
This composite span is just the span $(C\xleftarrow{K\circ P_Y}(J\downarrow G)\xrightarrow{H\circ P_X}A)$.

So in the 2-category $\mathcal{C}$, the composition of 1-morphisms $F^{\dag}\circ F$ is the following composition of spans,
\begin{equation}\label{fdf}
\begin{tikzpicture}[>=angle 60,thick,baseline=0pt]
  \node (a) at (0,1.6) {$(I\downarrow I)$};
  \node (b) at (-2,0) {$\mathbf{H}$};
  \node (c) at (2,0) {$\mathbf{H}$};
  \node (x) at (-4,-1.6) {$\mathbf{\Psi}$};
  \node (y) at (0,-1.6) {$\mathbf{\Psi}$};
  \node (z) at (4,-1.6) {$\mathbf{\Psi}$};
  \draw[->] (a) to node [above left] {$\pi_2$} (b);
  \draw[->] (a) to node [above right] {$\pi_1$} (c);
  \draw[->] (b) to node [above left] {$T$} (x);
  \draw[->] (b) to node [above right] {$I$} (y);
  \draw[->] (c) to node [above left] {$I$} (y);
  \draw[->] (c) to node [above right] {$T$} (z);
\end{tikzpicture}
\end{equation}
The groupoid $(I\downarrow I)$ has objects which are triples $(A,A,\alpha)$, and $A\xrightarrow{\alpha}A$ is an isomorphism in $\mathbf{\Psi}$. The projection maps $\pi_1$ and $\pi_2$ act in an obvious way on the objects.
The above composition can be rewritten as
\begin{equation}
\begin{tikzpicture}[>=angle 60,thick,baseline=0pt]
  \node (a) at (0,0.8) {$(I\downarrow I)$};
  \node (b) at (2,-.8) {$\mathbf{\Psi}$};
  \node (c) at (-2,-.8) {$\mathbf{\Psi}$};
  \draw[->] (a) to node [above left] {$T\circ\pi_2$} (c);
  \draw[->] (a) to node [above right] {$T\circ\pi_1$} (b);
\end{tikzpicture}
\end{equation}
This is just a 1-morphism $\mathbf{\Psi}\xrightarrow{F^{\dag}\circ F}\mathbf{\Psi}$ in the 2-category $\mathcal{C}$.

Similarly, the composition of spans $F\circ F^{\dag}$ is
\begin{equation}\label{ffd}
\begin{tikzpicture}[>=angle 60,thick,baseline=0pt]
  \node (a) at (0,1.6) {$(T\downarrow T)$};
   \node (b) at (-2,0) {$\mathbf{H}$};
  \node (c) at (2,0) {$\mathbf{H}$};
  \node (x) at (-4,-1.6) {$\mathbf{\Psi}$};
  \node (y) at (0,-1.6) {$\mathbf{\Psi}$};
  \node (z) at (4,-1.6) {$\mathbf{\Psi}$};
  \draw[->] (a) to node [above left] {$\pi_2$} (b);
  \draw[->] (a) to node [above right] {$\pi_1$} (c);
  \draw[->] (b) to node [above left] {$I$} (x);
  \draw[->] (b) to node [above right] {$T$} (y);
  \draw[->] (c) to node [above left] {$T$} (y);
  \draw[->] (c) to node [above right] {$I$} (z);
\end{tikzpicture}
\end{equation}
The groupoid $(T\downarrow T)$ has objects $(A,A,\beta)$, and $A^*\xrightarrow{\beta}A^*$ is an isomorphism in $\mathbf{\Psi}$.
The above composition can be rewritten as
\begin{equation}
\begin{tikzpicture}[>=angle 60,thick,baseline=0pt]
  \node (a) at (0,0.8) {$(T\downarrow T)$};
  \node (b) at (2,-.8) {$\mathbf{\Psi}$};
  \node (c) at (-2,-.8) {$\mathbf{\Psi}$};
  \draw[->] (a) to node [above left] {$I\circ\pi_2$} (c);
  \draw[->] (a) to node [above right] {$I\circ\pi_1$} (b);
\end{tikzpicture}
\end{equation}
This is a 1-morphism $\mathbf{\Psi}\xrightarrow{F\circ F^{\dag}}\mathbf{\Psi}$ in the 2-category $\mathcal{C}$.

It is easy to see that, there is an isomorphic relation $F\circ F^{\dag}\oplus F^{\dag}\circ F\cong\mathrm{id}_{\mathbf{\Psi}}$, this is just the relation (\ref{fa}).
There are no compositions of 1-morphisms like $F\circ F$ or $F^{\dag}\circ F^{\dag}$ in the 2-category $\mathcal{C}$. This is consistent with the results of the fermion algebra, since we have $\hat{f} \hat{f}=0$ and $\hat{f}^{\dag}\hat{f}^{\dag}=0$.

Now let us consider the 2-morphisms in $\mathcal{C}$, which are isomorphism classes of spans of spans.
A span of spans of type $(B\xleftarrow{G}X\xrightarrow{H}A)\to (B\xleftarrow{J}Y\xrightarrow{K}A)$ is a span $Y\xleftarrow{R}Z\xrightarrow{S}X$ equipped with natural isomorphisms $G\circ S\xrightarrow{\mu} J\circ R$ and $H\circ S\xrightarrow{\nu} K\circ R$, as indicated by the following diagram \cite{2Vec}
\begin{equation}
\begin{tikzpicture}[>=angle 60,thick,baseline=0pt]
  \node (a) at (0,-2) {$X$};
  \node (b) at (-2.5,0) {$B$};
  \node (c) at (2.5,0) {$A$};
  \node (y) at (0,2) {$Y$};
  \node (z) at (0,0) {$Z$};
  \draw[->] (a) to node [below left] {$G$} (b);
  \draw[->] (a) to node [below right] {$H$} (c);
  \draw[->] (y) to node [above left] {$J$} (b);
  \draw[->] (y) to node [above right] {$K$} (c);
  \draw[->] (z) to node [right] {$S$} (a);
  \draw[->] (z) to node [right] {$R$} (y);
  \draw[->,double distance=2pt,-implies] (-1.0,-.6) to node [right] {$\mu$} (-1.0,.6);
  \draw[->,double distance=2pt,-implies] (1.0,-.6) to node [right] {$\nu$} (1.0,.6);
\end{tikzpicture}
\end{equation}
We are considering such diagrams as 2-morphisms only up to isomorphism, namely, the inner span $Y\xleftarrow{R}Z\xrightarrow{S}X$ is only considered up to an isomorphism of spans in the sense of (\ref{iso}).

For example,
\begin{equation}\label{id1}
\begin{tikzpicture}[>=angle 60,thick,baseline=0pt]
  \node (a) at (0,2) {$(I\downarrow I)$};
  \node (b) at (-2.5,0) {$\mathbf{\Psi}$};
  \node (c) at (2.5,0) {$\mathbf{\Psi}$};
  \node (y) at (0,-2) {$\mathbf{\Psi}$};
  \node (z) at (0,0) {$\mathbf{H}$};
  \draw[->] (a) to node [above left] {$T\circ\pi_2$} (b);
  \draw[->] (a) to node [above right] {$T\circ\pi_1$} (c);
  \draw[->] (y) to node [below left] {$\mathrm{id}_{\mathbf{\Psi}}$} (b);
  \draw[->] (y) to node [below right] {$\mathrm{id}_{\mathbf{\Psi}}$} (c);
  \draw[->] (z) to node [right] {$\Delta_{I}$} (a);
  \draw[->] (z) to node [right] {$T$} (y);
  \draw[->,double distance=2pt,-implies] (-1.0,-.6) to node [right] {$\mathrm{id}$} (-1.0,.6);
  \draw[->,double distance=2pt,-implies] (1.0,-.6) to node [right] {$\mathrm{id}$} (1.0,.6);
\end{tikzpicture}
\end{equation}
where $\Delta_I$ is the diagonal functor, it takes objects $x\mapsto (x,x,\mathrm{id}_{I(x)})$ and morphisms $g\mapsto (g,g)$. This is a 2-morphism $\mathrm{id}_{\mathbf{\Psi}}\xrightarrow{\eta} F^{\dag}\circ F$ in $\mathcal{C}$.
It can also be represented by the following string diagram
\begin{equation}
\begin{tikzpicture}[>=angle 60,thick,baseline=0pt]
    \draw[<-] (-.7,.7) .. controls (-.7,-1) and (.7,-1) .. (.7,.7);
    \draw (0,-1.2) node {$\eta$};
    \useasboundingbox (-.8,-1.3) rectangle (.8,.5);
\end{tikzpicture}
\end{equation}
The 2-morphism $F\circ F^{\dag}\xrightarrow{\epsilon}\mathrm{id}_{\mathbf{\Psi}}$ corresponds to the following diagram
\begin{equation}\label{id0}
\begin{tikzpicture}[>=angle 60,thick,baseline=0pt]
  \node (a) at (0,-2) {$(T\downarrow T)$};
  \node (b) at (-2.5,0) {$\mathbf{\Psi}$};
  \node (c) at (2.5,0) {$\mathbf{\Psi}$};
  \node (y) at (0,2) {$\mathbf{\Psi}$};
  \node (z) at (0,0) {$\mathbf{H}$};
  \draw[->] (a) to node [below left] {$I\circ\pi_2$} (b);
  \draw[->] (a) to node [below right] {$I\circ\pi_1$} (c);
  \draw[->] (y) to node [above left] {$\mathrm{id}_{\mathbf{\Psi}}$} (b);
  \draw[->] (y) to node [above right] {$\mathrm{id}_{\mathbf{\Psi}}$} (c);
  \draw[->] (z) to node [right] {$\Delta_{T}$} (a);
  \draw[->] (z) to node [right] {$I$} (y);
  \draw[->,double distance=2pt,-implies] (-1.0,-.6) to node [right] {$\mathrm{id}$} (-1.0,.6);
  \draw[->,double distance=2pt,-implies] (1.0,-.6) to node [right] {$\mathrm{id}$} (1.0,.6);
\end{tikzpicture}
\end{equation}
This 2-morphism can be represented by the following string diagram
\begin{equation}
\begin{tikzpicture}[>=angle 60,thick,baseline=0pt]
    \draw[<-] (-.7,-.7) .. controls (-.7,1) and (.7,1) .. (.7,-.7);
    \draw (0,-1.3) node {$\epsilon$};
    \useasboundingbox (-.8,-1.3) rectangle (.8,.5);
\end{tikzpicture}
\end{equation}
We also have 2-morphisms corresponding to the following diagram
\begin{equation}
\begin{tikzpicture}[>=angle 60,thick,baseline=0pt]
    \draw[->] (-.7,-.6) .. controls (-.7,1.1) and (.7,1.1) .. (.7,-.6);
    \draw (0,-1.2) node {$\eta^{\dag}:F^{\dag}\circ F\to\mathrm{id}_{\mathbf{\Psi}}$};
\end{tikzpicture}\qquad\qquad\qquad\qquad
\begin{tikzpicture}[>=angle 60,thick,baseline=0pt]
    \draw[->] (-.7,.7) .. controls (-.7,-1) and (.7,-1) .. (.7,.7);
    \draw (0,-1.2) node {$\epsilon^{\dag}:\mathrm{id}_{\mathbf{\Psi}}\to F\circ F^{\dag}$};
\end{tikzpicture}
\end{equation}

It is easy to see that, there are no diagrams with crossings, because there are no any 2-morphisms like $F\circ F^{\dag}\to F^{\dag}\circ F$ or $F^{\dag}\circ F\to F\circ F^{\dag}$ in $\mathcal{C}$. This is consistent with the results of string diagrams in the previous section.

Using the constructions (\ref{id1}) and (\ref{id0}), one can verify the following relations by some straightforward calculations,
\begin{equation}
\mathrm{id}_{F\circ F^{\dag}}=\epsilon^{\dag}\circ\epsilon,\qquad
\mathrm{id}_{F^{\dag}\circ F}=\eta\circ\eta^{\dag},\qquad
\mathrm{id}_{\mathrm{id}_{\mathbf{\Psi}}}
=\epsilon\circ\epsilon^{\dag}+\eta^{\dag}\circ \eta.
\end{equation}
Note that the 2-morphisms in the 2-category $\mathcal{C}$ are isomorphism classes of spans of spans, so these spans of spans are equal in the sense of isomorphism.
The first relation is equivalent to the first diagrammatic relation in (\ref{lcr1}), the second relation is equivalent to the first diagrammatic relation in (\ref{lcr3}), and the last relation is equivalent to the diagrammatic relation (\ref{lcr2}).

Using the results in Ref.~\cite{CHA}, one may find that there is an ambidextrous adjunction between the spans $F$ and $F^{\dag}$, and the 2-morphisms $\eta$, $\epsilon^{\dag}$ and $\epsilon$, $\eta^{\dag}$ are the corresponding units and counits. So we have the following adjunction equations
\begin{eqnarray}
&&(\eta^{\dag}\circ\mathrm{id}_{F^{\dag}})\ast(\mathrm{id}_{F^{\dag}}\circ\epsilon^{\dag})
=\mathrm{id}_{F^{\dag}}
=(\mathrm{id}_{F^{\dag}}\circ\eta)\ast(\epsilon\circ\mathrm{id}_{F^{\dag}}),
\nonumber\\
&&~~~~~(\epsilon\circ\mathrm{id}_{F})\ast(\mathrm{id}_{F}\circ\eta)
=\mathrm{id}_{F}
=(\mathrm{id}_{F}\circ\epsilon^{\dag})\ast(\eta^{\dag}\circ\mathrm{id}_{F}),
\end{eqnarray}
here ``$\circ$'' and ``$\ast$'' denote the vertical and horizontal composition of 2-morphisms, respectively.
These are just the zig-zag rules in the string diagrams
\begin{equation}
\begin{tikzpicture}[>=angle 60,thick,baseline=0pt]
  \draw (0,0) .. controls (0,1) and (.75,1) .. (.75,0);
  \draw (0,-1) -- (0,0);
  \draw (1.5,0) .. controls (1.5,-1) and (.75,-1) .. (.75,0);
  \draw (1.5,0) -- (1.5,1) [->];
  \draw (2.5,0) node {$=$};
  \draw (3.5,-1) -- (3.5,1) [->];
  \draw (4.5,0) node {$=$};
  \draw (5.5,0) -- (5.5,1) [->];
  \draw (5.5,0) .. controls (5.5,-1) and (6.25,-1) .. (6.25,0);
  \draw (6.25,0) .. controls (6.25,1) and (7,1) .. (7,0);
  \draw (7,-1) -- (7,0);
  \useasboundingbox (-.1,-1.1) rectangle (7.1,1.1);
  \end{tikzpicture}
\end{equation}
\begin{equation}
\begin{tikzpicture}[>=angle 60,thick,baseline=0pt]
  \draw (0,0) .. controls (0,1) and (.75,1) .. (.75,0);
  \draw (0,-1) -- (0,0) [<-];
  \draw (1.5,0) .. controls (1.5,-1) and (.75,-1) .. (.75,0);
  \draw (1.5,0) -- (1.5,1);
  \draw (2.5,0) node {$=$};
  \draw (3.5,-1) -- (3.5,1) [<-];
  \draw (4.5,0) node {$=$};
  \draw (5.5,0) -- (5.5,1);
  \draw (5.5,0) .. controls (5.5,-1) and (6.25,-1) .. (6.25,0);
  \draw (6.25,0) .. controls (6.25,1) and (7,1) .. (7,0);
  \draw (7,-1) -- (7,0) [<-];
  \useasboundingbox (-.1,-1.1) rectangle (7.1,1.1);
  \end{tikzpicture}
\end{equation}
\ \\
This is the isotopy condition of strands in the graphical category.

\section{Conclusions and discussions}\label{sec5}
In this paper, based on the methods developed in \cite{Groupoid, CHA}, we studied the groupoidification of the fermion algebra. We constructed groupoids corresponding to the fermionic Fock space in an intuitive way, and the fermionic creation and annihilation operators correspond to some types of spans of groupoids.
We find that the construction of spans for the fermionic creation and annihilation operators are a little different from those in the bosonic case in Ref.~\cite{CHA}.
We also construct a 2-category of spans of groupoids, and we found that the relations of the 2-morphisms are consistent with those in the graphical category constructed in Ref.~\cite{lwwy}.

Since the fermion algebraic relations and the corresponding Fock space are much simpler than those of the boson algebra, we found that the groupoidification of the fermion algebra is also much simpler than that of the boson algebra. This is consistent with the results of the diagrammatic categorification with the methods of string diagrams.

The fermionic Fock space corresponds to the groupoid $\mathbf{\Psi}$, and the Fock states $|0\rangle$ and $|1\rangle$ correspond to the object $A$ and its dual object $A^*$ in $\mathbf{\Psi}$. If $\mathrm{Aut}(A)$ and $\mathrm{Aut}(A^*)$ are trivial groups, namely $\mathrm{Hom}_{\mathbf{\Psi}}(A,A)=\{\mathrm{id}_A\}$ and $\mathrm{Hom}_{\mathbf{\Psi}}(A^*,A^*)=\{\mathrm{id}_{A^*}\}$, then the groupoid $\mathbf{\Psi}$ is just a discrete groupoid, in fact, a set. In this case, the categorical analogues of the fermionic Fock states $|0\rangle$ and $|1\rangle$ have no any extra structures. This is a trivial case of categorification. If $\mathrm{Aut}(A)$ and $\mathrm{Aut}(A^*)$ are notrivial groups, then there are some additional structures on the categorical fermionic Fock states. In this case, the categorical fermionic Fock states may have some additional properties, and these additional properties maybe depend on the concrete physical systems. Different additional structures maybe correspond to different fermion systems. So one may use these categorical fermionic Fock states to describe
the fermion systems more finely, and study some additional properties of the fermion systems.

Our methods can be easily extended to the study of higher-dimensional fermion algebras.
The present work also provides new insight into the fermion algebra and the corresponding quantum physics.
Since the boson and fermion algebras are the simplest and most fundamental algebraic relations in quantum physics, our result is a complement to the study of groupoidification and categorification of physical theories.

\section*{Acknowledgements}
This project is supported by the National Natural Science Foundation of China (Nos. 11405060, 11571119).

\end{document}